\documentstyle[preprint,aps,floats,psfig,epsf]{revtex}
\catcode`@=11
\def\references{%
\ifpreprintsty
\bigskip\bigskip
\hbox to\hsize{\hss\large \refname\hss}%
\else
\vskip24pt
\hrule width\hsize\relax
\vskip 1.6cm
\fi
\list{\@biblabel{\arabic{enumiv}}}%
{\labelwidth\WidestRefLabelThusFar  \labelsep4pt %
\leftmargin\labelwidth %
\advance\leftmargin\labelsep %
\ifdim\baselinestretch pt>1 pt %
\parsep  4pt\relax %
\else %
\parsep  0pt\relax %
\fi
\itemsep\parsep %
\usecounter{enumiv}%
\let\p@enumiv\@empty
\def\theenumiv{\arabic{enumiv}}%
}%
\let\newblock\relax %
\sloppy\clubpenalty4000\widowpenalty4000
\sfcode`\.=1000\relax
\ifpreprintsty\else\small\fi
}
\catcode`@=12

\def\lsim{\mathrel{\raise.3ex\hbox{$<$\kern-.75em\lower1ex\hbox{$\sim$}}}}
\def\gsim{\mathrel{\raise.3ex\hbox{$>$\kern-.75em\lower1ex\hbox{$\sim$}}}}
\begin{document}
\tightenlines

\hfill\vtop{
\hbox{AMES-HET-02-07}
\hbox{BUHEP-02-37}
\hbox{MAD-PH-1311}
\hbox{hep-ph/0210428}
\hbox{}}

\vspace*{.25in}
\begin{center}
{\large\bf How two neutrino superbeam experiments\\
do better than one}\\[10mm]
V. Barger$^1$, D. Marfatia$^2$ and K. Whisnant$^3$\\[5mm]
\it
$^1$Department of Physics, University of Wisconsin,
Madison, WI 53706, USA\\
$^2$Department of Physics, Boston University,
Boston, MA 02215, USA\\
$^3$Department of Physics and Astronomy, Iowa State University,
Ames, IA 50011, USA

\end{center}
\thispagestyle{empty}

\begin{abstract}

\vspace*{-.35in}

\noindent

We examine the use of two superbeam neutrino oscillation experiments
with baselines $\lsim 1000$~km to resolve parameter degeneracies
inherent in the three-neutrino analysis of such experiments. We find
that with appropriate choices of neutrino energies and baselines two
experiments with different baselines can provide a much better
determination of the neutrino mass ordering than a single experiment
alone. Two baselines are especially beneficial when the mass scale for
solar neutrino oscillations $\delta m^2_{\rm sol}$ is $\gsim
5\times10^{-5}$~eV$^2$. We also examine $CP$ violation sensitivity and
the resolution of other parameter degeneracies. We find that the
combined data of superbeam experiments with baselines of 295 and
900~km can provide sensitivity to both the neutrino mass ordering and
$CP$ violation for $\sin^22\theta_{13}$ down to 0.03 for $|\delta
m^2_{\rm atm}| \simeq 3\times10^{-3}$~eV$^2$. It would be advantageous
to have a 10\% determination of $|\delta m^2_{\rm atm}|$ before the
beam energies and baselines are finalized, although if $|\delta m^2_{\rm
atm}|$ is not that well known, the neutrino energies and baselines can be
chosen to give fairly good sensitivity for a range of $|\delta
m^2_{\rm atm}|$.

\end{abstract}

\newpage

\section{Introduction}

Atmospheric neutrino data from Super-Kamiokande provides strong evidence
that $\nu_\mu$'s created in the atmosphere oscillate to $\nu_\tau$
with mass-squared difference $|\delta m^2_{\rm atm}| \sim 3 \times
10^{-3}$~eV$^2$ and almost maximal amplitude~\cite{Toshito:2001dk}.
Furthermore, the recent solar neutrino data from the Sudbury Neutrino
Observatory (SNO) establishes that electron neutrinos change flavor as
they travel from the Sun to the Earth: the neutral-current measurement
is consistent with the solar neutrino flux predicted in the Standard
Solar Model~\cite{Bahcall:2000nu}, while the charged-current
measurement shows a depletion of the electron neutrino component
relative to the total flux~\cite{SNONC}. Global fits to solar neutrino
data give a strong preference for the Large Mixing Angle (LMA)
solution to the solar neutrino puzzle, with $\delta m^2_{\rm sol} \sim
5 \times 10^{-5}$ eV$^2$ and amplitude close to 0.8~\cite{SNONC,LMA}.


The combined atmospheric and solar data may be explained by
oscillations of three neutrinos, that are described by two
mass-squared differences, three mixing angles and a $CP$ violating
phase. The atmospheric and solar data roughly determine $\delta
m^2_{\rm atm}$, $\delta m^2_{\rm sol}$ and the corresponding mixing
angles. The LMA solar solution will be tested decisively (and $\delta
m^2_{\rm sol}$ measured accurately) by the KamLAND reactor neutrino
experiment~\cite{kamland,bmwood}. More precise measurements of the
other oscillation parameters may be performed in long-baseline
neutrino experiments. The low energy beam at MINOS~\cite{minos} plus
experiments with ICARUS~\cite{icarus} and OPERA~\cite{opera} will
allow an accurate determination of the atmospheric neutrino parameters
and may provide the first evidence for oscillations of $\nu_\mu \to
\nu_e$ at the atmospheric mass scale~\cite{bgmtw}. It will take a new
generation of long-baseline experiments to further probe $\nu_\mu \to
\nu_e$ appearance and to measure the leptonic $CP$ phase. Matter
effects are the only means to determine sgn($\delta m^2_{\rm atm}$);
once sgn($\delta m^2_{\rm atm}$) is known, the level of intrinsic $CP$
violation may be measured. Matter effects and intrinsic $CP$ violation
both vanish in the limit that the mixing angle responsible for
$\nu_\mu \to \nu_e$ oscillations of atmospheric neutrinos is zero.

It is now well-known that there are three two-fold parameter
degeneracies that can occur in the measurement of the oscillation
amplitude for $\nu_\mu \to \nu_e$ appearance, the ordering of the
neutrino masses, and the $CP$ phase~\cite{bmw}. With only one $\nu$
and one $\bar\nu$ measurement, these degeneracies can lead to eight
possible solutions for the oscillation parameters; in most cases, $CP$
violating ($CPV$) and $CP$ conserving ($CPC$) solutions can equally
explain the same data. Studies have been done on how a
superbeam~\cite{bmw,bmw2,sign,minakata2,barenboim,huber},
neutrino factory~\cite{huber,ambiguity,nufacts}, superbeam plus
neutrino factory~\cite{mena}, or two superbeams with one at a very
long baseline~\cite{VLBL,hagiwara} could be used to resolve one or
more of these ambiguities.

In this paper we show that by combining the results of two superbeam
experiments with different medium baselines, $\lsim 1000$~km, the
ambiguity associated with the sign of $\delta m^2_{\rm atm}$ can be
resolved, even when it cannot be resolved by the two experiments taken
separately. Furthermore, the ability to determine sgn($\delta m^2_{\rm
atm}$) from the combined data is found to not be greatly sensitive to
the size of $\delta m^2_{\rm sol}$, unlike the situation where data
from only a single baseline is used. If both experiments are at or
near the peak of the oscillation, a good compromise is obtained
between the sensitivities for resolving sgn($\delta m^2_{\rm atm}$)
and for establishing the existence of $CP$ violation. If $|\delta
m^2_{\rm atm}|$ is not known accurately, the neutrino energies and
baselines can be chosen to give fairly good sensitivity to the sign of
$\delta m^2_{\rm atm}$ and to $CP$ violation for a range of $|\delta
m^2_{\rm atm}|$.

The organization of our paper is as follows. In Sec.~\ref{sec:degen}
we discuss the parameter degeneracies that can occur in the analysis
of long-baseline oscillation data. In Sec.~\ref{sec:joint} we analyze
how two long-baseline superbeam experiments can break degeneracies,
determine the neutrino parameters, and establish the existence of $CP$
violation in the neutrino sector, if it is present. A summary is
presented in Sec.~\ref{sec:summary}.

\section{Parameter degeneracies}
\label{sec:degen}

We work in the three-neutrino scenario using the parametrization for
the neutrino mixing matrix of Ref.~\cite{bmw}. If we assume that
$\nu_3$ is the neutrino eigenstate that is separated from the other
two, then $\delta m^2_{31} = \delta m^2_{\rm atm}$ and the sign of
$\delta m^2_{31}$ can be either positive or negative, corresponding to
the mass of $\nu_3$ being either larger or smaller, respectively, than
the other two masses. The solar oscillations are regulated
by $\delta m^2_{21} = \delta m^2_{\rm sol}$, and thus $|\delta
m^2_{21}| \ll |\delta m^2_{31}|$. If we accept the likely conclusion
that the solar solution is LMA~\cite{SNONC,LMA}, then $\delta m^2_{21}
> 0$ and we can restrict $\theta_{12}$ to the range $[0,\pi/4]$.  It
is known from reactor neutrino data that $\theta_{13}$ is small, with
$\sin^22\theta_{13} \le 0.1$ at the 95\% C.L.~\cite{CHOOZ}. Thus a set
of parameters that unambiguously spans the space is $\delta m^2_{31}$
(magnitude and sign), $\delta m^2_{21}$, $\sin^22\theta_{12}$,
$\sin\theta_{23}$, and $\sin^22\theta_{13}$; only the $\theta_{23}$
angle can be below or above $\pi/4$. 

For the oscillation probabilities for $\nu_\mu \to \nu_e$ and
$\bar\nu_\mu \to \bar\nu_e$ we use approximate expressions given
in Ref.~\cite{bmw}, in which the probabilities are expanded in terms
of the small parameters $\theta_{13}$ and $\delta
m^2_{21}$~\cite{cervera,freund}, which reproduces well the exact
oscillation probabilities for $E_\nu \gsim 0.5$~GeV, $\theta_{13} \lsim
9^\circ$, and $L \lsim 4000$~km~\cite{bmw}. In all of our calculations we
use the average electron density along the neutrino path,
assuming the Preliminary Reference Earth Model~\cite{PREM}. Our
calculational methods are described in Ref.~\cite{bmw2}.

We expect that $|\delta m^2_{31}|$ and $\sin^22\theta_{23}$ will be
measured to an accuracy of $\simeq 10$\% at $3\sigma$ from $\nu_\mu
\to \nu_\mu$ survival in long-baseline
experiments~\cite{minos,icarus,opera,bgmtw}, while $\delta m^2_{21}$
will be measured to an accuracy of $\simeq 10$\% at $2\sigma$ and
$\sin^22\theta_{12}$ will be measured to an accuracy of $\pm0.1$ at
$2\sigma$ in experiments with reactor neutrinos~\cite{bmwood}. The
remaining parameters ($\theta_{13}$, the $CP$ phase $\delta$, and the
sign of $\delta m^2_{31}$) must be determined from long-baseline
appearance experiments, principally using the modes $\nu_\mu \to
\nu_e$ and $\bar\nu_\mu \to \bar\nu_e$ with conventional neutrino
beams, or $\nu_e \to \nu_\mu$ and $\bar\nu_e \to \bar\nu_\mu$ at
neutrino factories.  However, there are three parameter degeneracies
that can occur in such an analysis: (i) the ($\delta,\theta_{13}$)
ambiguity~\cite{ambiguity}, (ii) the sgn($\delta m^2_{31}$)
ambiguity~\cite{sign}, and (iii) the ($\theta_{23},\pi/2 -
\theta_{23}$) ambiguity~\cite{bmw} (see Ref.~\cite{bmw} for a complete
discussion of these three parameter degeneracies). In each degeneracy,
two different sets of values for $\delta$ and $\theta_{13}$ can give
the same measured rates for both $\nu$ and $\bar\nu$ appearance and
disappearance. For each type of degeneracy the values of $\theta_{13}$
for the two equivalent solutions can be quite different, and the two
values of $\delta$ may have different $CP$ properties, e.g., one can
be $CP$ conserving and the other $CP$ violating.

A judicious choice of $L$ and $E_\nu$ can reduce the impact of the
degeneracies. For example, if $L/E_\nu$ is chosen such that $\Delta
\equiv |\delta m^2_{31}| L/(4E_\nu) = \pi/2$ (the peak of the
oscillation in vacuum), then the $\cos\delta$ terms in the average
appearance probabilities vanish, even after matter effects are
included~\cite{bmw}. Then since it is $\sin\delta$ that is being
measured, the ($\delta,\theta_{13}$) ambiguity is reduced to a simple
($\delta, \pi - \delta$) ambiguity, $CPV$ solutions are no longer
mixed with $CPC$ solutions, and $\theta_{13}$ is in principle
determined (for a given sgn($\delta m^2_{31}$) and $\theta_{23}$). If
$L$ is chosen to be long enough ($\gsim 1000$~km), then the
predictions for $\delta m^2_{31}>0$ and $\delta m^2_{31}<0$ no longer
overlap if $\theta_{13} \gsim$ a few degrees, and the sgn($\delta
m^2_{31}$) ambiguity is removed; our previous studies indicated that
for $\delta m^2_{21} = 5\times10^{-5}$~eV$^2$ this happens at $L \gsim
1300$~km if $\sin^22\theta_{13} > 0.01$~\cite{bmw} (before
experimental uncertainties are considered). However, the persistence
of the sgn($\delta m^2_{31}$) ambiguity is highly dependent on the
size of the solar oscillation mass scale, because large values of
$\delta m^2_{21}$ cause the predictions for $\delta m^2_{31}>0$ and
$\delta m^2_{31}<0$ to overlap much more severely than when $\delta
m^2_{21}$ is smaller. Also, existing neutrino baselines are no longer
than 735~km. In this paper we explore the possibility that two
experiments with medium baselines ($\lsim 1000$~km) can determine
sgn($\delta m^2_{31}$), even when data from one of the baselines
alone cannot. We then address the sensitivity for establishing $CP$
violation.

\section{Joint analysis of two superbeam experiments}
\label{sec:joint}

\subsection{Description of the experiments and method}
\label{sec:experiments}

For our analysis we take one baseline to be 295~km, the distance for the
proposed experiment from the Japan Hadron facility (JHF) to the
Super-Kamiokande detector at Kamioka. For the neutrino spectrum of
this experiment we use their $2^\circ$ off-axis beam with average
neutrino energy of 0.7~GeV~\cite{JHFSKspectra}. For the second
experiment we assume an off-axis beam in which the beam axis points
at a site 735~km from the source (appropriate for a beamline from NuMI
at Fermilab to Soudan, or from CERN to Gran Sasso). For the off-axis
spectra of the NuMI experiment we use the results presented in
Ref.~\cite{OAspectra}, which provides neutrino spectra for 39 different
off-axis angles ranging from $0.32^\circ$ to $1.76^\circ$.

Using the off-axis components of the beam has the advantage of a lower
background~\cite{barenboim,bnl,para} due to reduced $\nu_e$
contamination and a smaller high-energy tail. Off-axis beams also
offer flexibility in the choice of $L$ and $E_\nu$. For example, for a
beam nominally aimed at a ground-level site a distance $L_0$ from the
source, the distance to a ground-level detector with off-axis angle
$\theta_{OA}$ can lie anywhere in the range
\begin{equation}
2 R_e \sin(\theta - \theta_{OA}) \le L \le
2 R_e \sin(\theta + \theta_{OA}) \,,
\label{eq:LOA1}
\end{equation}
where $\sin\theta = L_0/(2 R_e)$, and $R_e = 6371$~km is the radius of the
Earth. Then for $L_0^2 \ll R_e^2$ the possible range of distances for an
off-axis detector at approximately ground level is
\begin{equation}
L_0 - 2 R_e \theta_{OA} \lsim L \lsim L_0 + 2 R_e \theta_{OA} \,.
\label{eq:LOA2}
\end{equation}
The neutrino energy and neutrino flux $\Phi_\nu$ decrease with increasing
off-axis angle as
\begin{equation}
E_\nu = {0.43 E_\pi\over 1 + \gamma^2 \theta_{OA}^2} ,~~~~~
\Phi_\nu \propto {E_\nu^2 \over L^2} \,,
\label{enuphi}
\end{equation}
where $\gamma = E_\pi/m_\pi$ is boost factor of the decaying pion.
Thus a wide range of $L$ and $E_\nu$ can be achieved with a single
fixed beam, although the event rate will drop with increasing
off-axis angle because the flux decreases and the neutrino cross
section is smaller at smaller $E_\nu$ (thereby putting a limit on the
usable range of $L$ and $E_\nu$).
 

For the first experiment at $L_1 = 295$~km, we assume that the
neutrino spectrum is chosen so that the $\cos\delta$ terms in the
$\nu$ and $\bar\nu$ oscillation probabilities vanish (after averaging
over the neutrino spectrum), using the best existing experimental
value for $\delta m^2_{31}$. The JHF $2^\circ$ off-axis
beam~\cite{jhfsk} satisfies this condition for $\delta m^2_{31} =
3\times10^{-3}$~eV$^2$. This spectrum choice reduces the ($\delta,
\theta_{13}$) ambiguity to a simple ($\delta, \pi-\delta$) ambiguity,
as described in Sec.~\ref{sec:degen}. For the second experiment we
allow $L_2$ and $\theta_{OA}$ to vary within the restrictions of
Eq.~(\ref{eq:LOA2}). This flexibility can be fully utilized if a deep
underground site is not required; the short duration of the beam
operation (an 8.6~$\mu$s pulse with a 1.9~s cycle time~\cite{NuMI})
may enable a sufficient reduction in the cosmic ray neutrino
background. We assume that the proton drivers at the neutrino sources
have been upgraded from their initial designs (from 0.8 to 4.0~MW for
JHF~\cite{jhfsk} and from 0.4 to 1.6~MW for FNAL~\cite{NuMIupgrade}),
so that they are both true neutrino superbeams. We assume two years
running with neutrinos and six years with antineutrinos at JHF, and
two years with neutrinos and five years with antineutrinos at FNAL;
these running times give approximately equivalent numbers of
charged-current events for neutrinos and antineutrinos at the two
facilities, in the absence of oscillations. For detectors, we assume a
22.5~kt detector in the JHF beam (such as the current Super-K
detector) and a 20~kt detector in the FNAL beam (which was proposed in
Ref.~\cite{barenboim}). Larger detectors such as Hyper-Kamiokande or
UNO would allow shorter beam exposures or higher precision studies. In
all of our calculations, we assume $|\delta m^2_{31}| =
3\times10^{-3}$~eV$^2$, $\theta_{23} = \pi/4$, $\delta m^2_{21} =
5\times10^{-5}$~eV$^2$, and $\sin^22\theta_{12} = 0.8$, unless noted
otherwise.

\begin{figure}
\centering\leavevmode
\mbox{\psfig{file=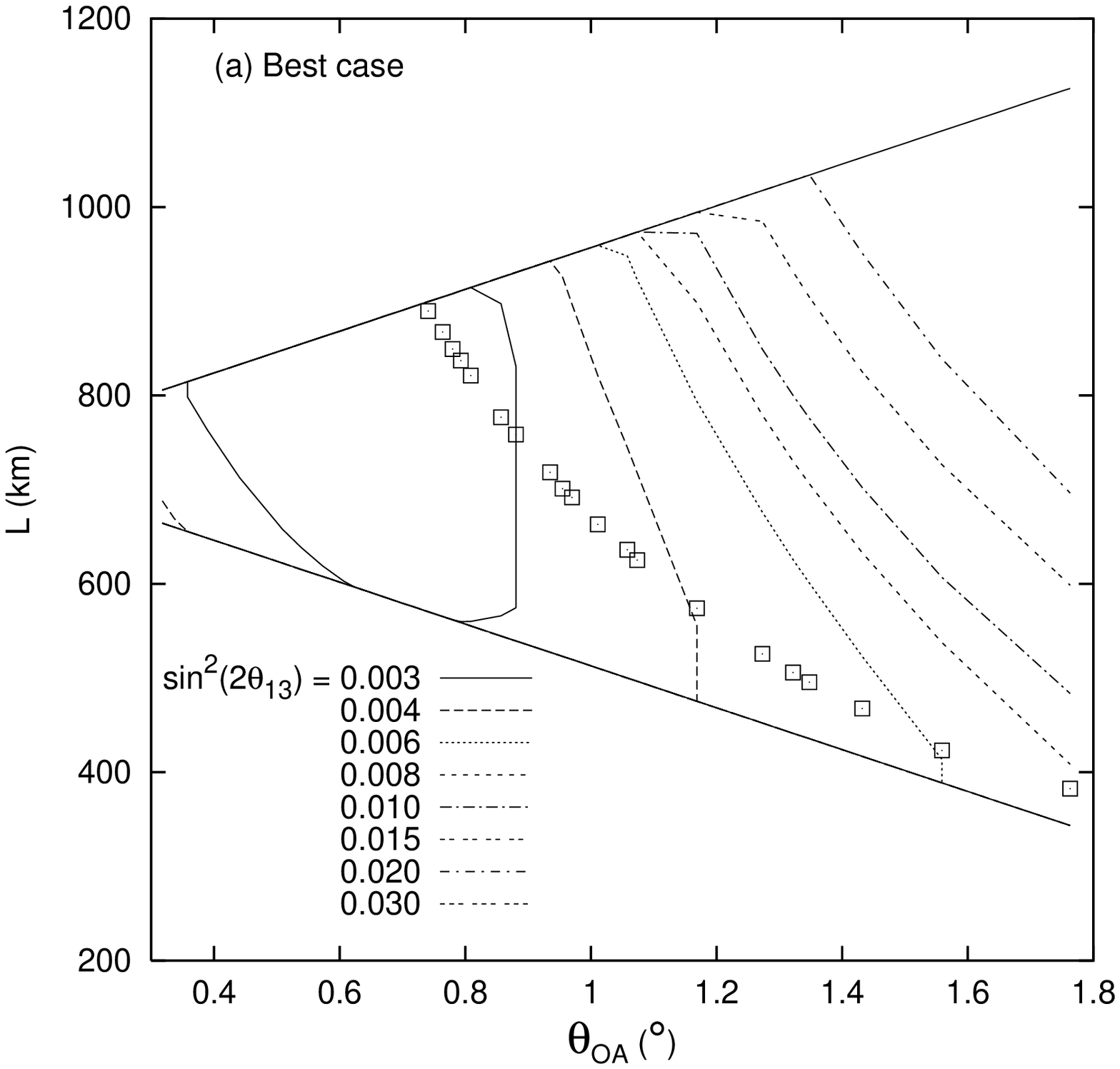,width=8cm,height=12cm}
\psfig{file=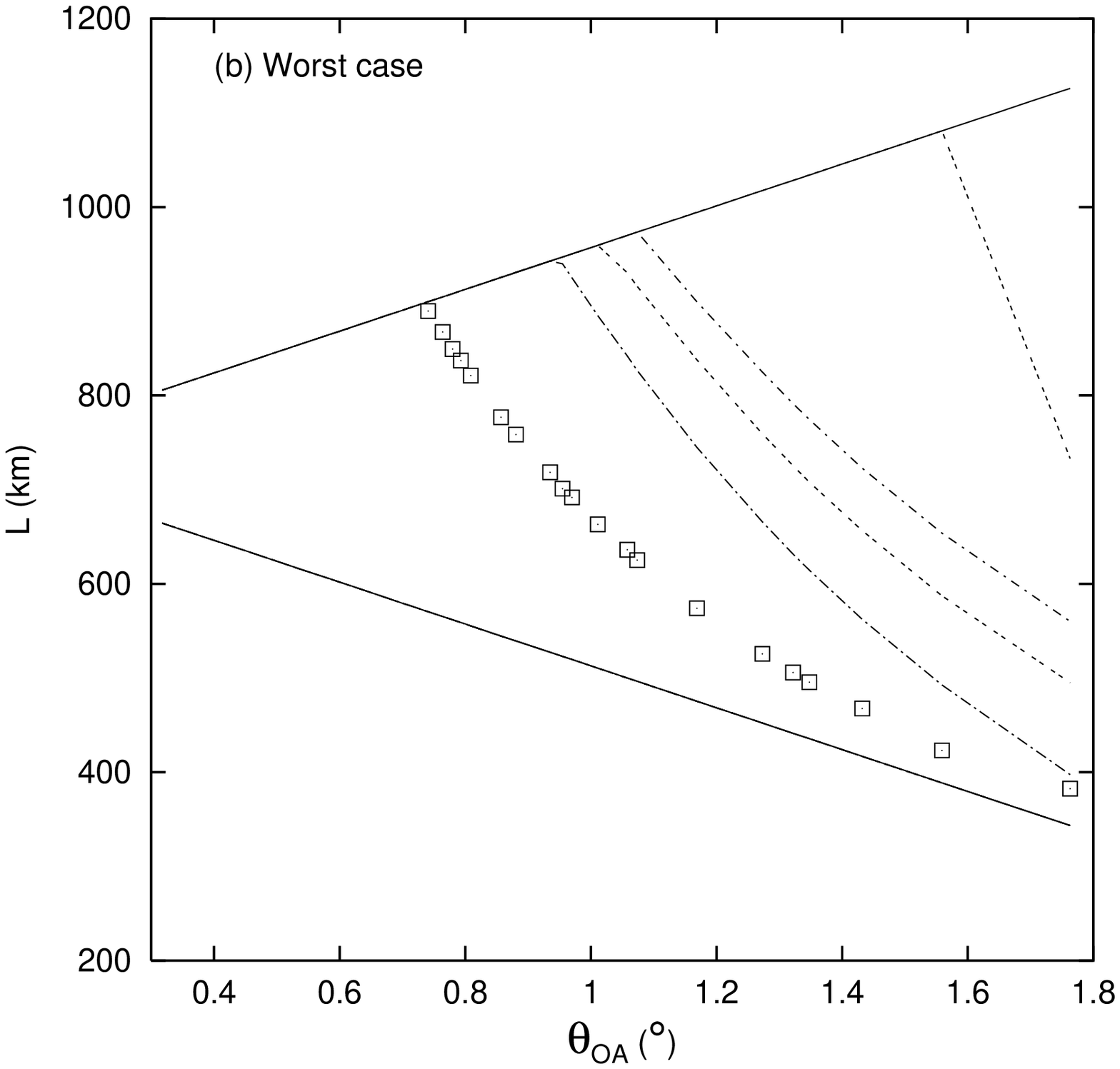,width=8cm,height=12cm}}
\medskip
\caption{Contours of (a) best-case (when $\delta m^2_{31} > 0$), and
(b) worst-case (when $\delta m^2_{31} < 0$), $\sin^22\theta_{13}$
$3\sigma$ discovery reach in the ($\theta_{OA}, L_2$) plane, for the
$\nu$ channel at NuMI, where $\theta_{OA}$ is the off-axis angle and
$L_2$ is the baseline of the NuMI detector. For the other neutrino
parameters we assume $|\delta m^2_{31}| = 3\times10^{-3}$~eV$^2$,
$\theta_{23} = \pi/4$, $\delta m^2_{21} = 5\times10^{-5}$~eV$^2$, and
$\sin^22\theta_{12} = 0.8$. The boxes indicate detector positions for
which the $\cos\delta$ terms in the average oscillation probabilities
vanish. For the $\bar\nu$ channel the results are similar, except that
the best case occurs for $\delta m^2_{31} < 0$ and the worst case for
$\delta m^2_{31} > 0$.}
\label{fig:discovery}
\end{figure}

We first consider the minimum value of $\sin^22\theta_{13}$ for which
the signal in the neutrino appearance channel can be seen above
background at the $3\sigma$ level (the discovery reach), varying over
a range of allowed values for $\theta_{OA}$ and $L_2$ in the second
experiment. The discovery reach depends on the value of $\delta$ and
the sign of $\delta m^2_{31}$; the best (when $\delta m^2_{31} > 0$)
and worst (when $\delta m^2_{31} < 0$) cases in the $\nu$ channel
(after varying over $\delta$) are shown in Fig.~\ref{fig:discovery}.
In the $\bar\nu$ channel, the best case occurs for $\delta m^2_{31} <
0$ and the worst for $\delta m^2_{31} > 0$. In our calculations we
assume a background that is 0.5\% of the unoscillated charged-current
rate (see Ref.~\cite{barenboim}), and that the systematic error is 5\%
of the background. However, we note that our general conclusions are
not significantly affected by reasonable changes in these experimental
uncertainty assumptions. Detector positions where there is no
$\cos\delta$ dependence in the rates are denoted by boxes. The best
reach is $\sin^22\theta_{13} \simeq 0.003$, which occurs for
$\theta_{OA} \simeq 0.5$-$0.9^\circ$. In the worst case scenario the
reach degrades to $\sin^22\theta_{13} \simeq 0.01$.

The measurement of $P$ and $\bar P$ at $L_1$ allows a determination of
$\sin^22\theta_{13}$ and $\sin\delta$, modulo the possible uncertainty
caused by the sign of $\delta m^2_{31}$, assuming for the moment that
$\theta_{23} = \pi/4$, so there is no ($\theta_{23}, \pi/2 -
\theta_{23}$) ambiguity.  The question we next consider is whether
an additional measurement of $P$ and $\bar P$ at $L_2$ can
determine sgn($\delta m^2_{31}$), measure $CP$ violation, and
distinguish $\delta$ from $\pi - \delta$. We define the $\chi^2$ of
neutrino parameters ($\delta^\prime, \theta_{13}^\prime$) relative to
the parameters ($\delta,\theta_{13}$) as
\begin{equation}
\chi^2 = \sum_i {(N_i - N_i^\prime)^2 \over (\delta N_i)^2} \,,
\label{eq:chisq}
\end{equation}
where $N_i$ and $N_i^\prime$ are the event rates for the parameters
($\delta,\theta_{13}$) and ($\delta^\prime,\theta_{13}^\prime$),
respectively, $\delta N_i$ is the uncertainty in $N_i$, and $i$ is
summed over the measurements being used in the analysis ($\nu$ and
$\bar\nu$ at $L_1$ and $\nu$ and $\bar\nu$ at $L_2$). For $\delta N_i$
we assume that the statistical error for the signal plus background can
be added in quadrature with the systematic error. For a two-parameter
system ($\delta$ and $\theta_{13}$ unknown), two sets of parameters
can be resolved at the $2\sigma$ ($3\sigma$) level if $\chi^2 > 6.17$
($11.83$).

\subsection{Determining the sign of $\delta m^2_{31}$}
\label{sec:sign}


To determine if measurements at $L_1$ and $L_2$ can distinguish one
set of oscillation parameters with one sign of $\delta m^2_{31}$ from
all other possible sets of oscillation parameters with the opposite
sign of $\delta m^2_{31}$, we sample the ($\delta,\theta_{13}$) space
for the opposite sgn($\delta m^2_{31}$) using a fine grid with
$1^\circ$ spacing in $\delta$ and approximately 2\% increments in
$\sin^22\theta_{13}$. If the $\chi^2$ between the original set of
oscillation parameters and all of those with the opposite sgn($\delta
m^2_{31}$) is greater than $6.17$ ($11.83$), then sgn($\delta
m^2_{31}$) is distinguished at the $2\sigma$ ($3\sigma$) level for
that parameter set.


Figure~\ref{fig:sign2} shows contours (in the space of possible $L_2$
and $\theta_{OA}$ for the second experiment) for the minimum value of
$\sin^22\theta_{13}$ (the $\sin^22\theta_{13}$ reach) for
distinguishing sgn($\delta m^2_{31}$) at the $3\sigma$ level when
$\nu$ and $\bar\nu$ data from $L_1$ and $L_2$ are combined. As in
Fig.~\ref{fig:discovery}, the boxes indicate the detector positions
where the $\cos\delta$ terms in the average probabilities vanish. The
best reach of about $\sin^22\theta_{13} \simeq 0.03$ can be realized
for $\theta_{OA} \simeq 0.7$-$1.0^\circ$ and $L_2$ values near the
maximum allowed by Eq.~\ref{eq:LOA2} ($\simeq 875$-$950$~km).
Table~\ref{tab:sign1} shows the sensitivity for determining
sgn($\delta m^2_{31}$) for different combinations of detector size and
proton driver power in the two experiments. The table shows that once
enough statistics are obtained at JHF (with a 22.5~kt detector and a
4~MW source), combined JHF and NuMI data significantly improve the
$\sin^22\theta_{13}$ reach for determining sgn($\delta m^2_{31}$) at
$3\sigma$ (by nearly a factor of two compared to data from a 1.6~MW
NuMI alone).

\begin{table}[hbtp!]
\begin{center}
\caption{$\sin^22\theta_{13}$ reach for determining the sign of $\delta
m^2_{31}$ at $3\sigma$ using $\nu$ and $\bar\nu$ data from JHF at 295~km
and NuMI at $L_2$, for various detector sizes and proton driver powers.
The approximate range of $\theta_{OA}$ that can obtain the reach shown is
given in parentheses; $L_2 \sim 900$~km in all cases.} 
\begin{tabular}{c|cc}
                & \multicolumn{2}{c}{NuMI (20 kt)}\\
JHF             & 0.4~MW                  & 1.6~MW\\
                & $\sin^22\theta_{13}$ ($\theta_{OA}$)
& $\sin^22\theta_{13}$ ($\theta_{OA}$)\\
\hline\hline
no JHF data     & 0.09 (0.7-1.0$^\circ$) & 0.05 (0.8-1.0$^\circ$)\\
\hline
22.5~kt, 0.8~MW & 0.07 (0.8-1.0$^\circ$) & 0.04 (0.9-1.0$^\circ$)\\
22.5~kt, 4.0~MW & 0.06 (0.7-1.0$^\circ$) & 0.03 (0.7-1.0$^\circ$)\\
 450~kt, 4.0~MW & 0.05 (0.6-1.0$^\circ$) & 0.02 (0.7-0.9$^\circ$)\\
\end{tabular}
\label{tab:sign1}
\end{center}
\end{table}

\begin{figure}
\centering\leavevmode
\mbox{\psfig{file=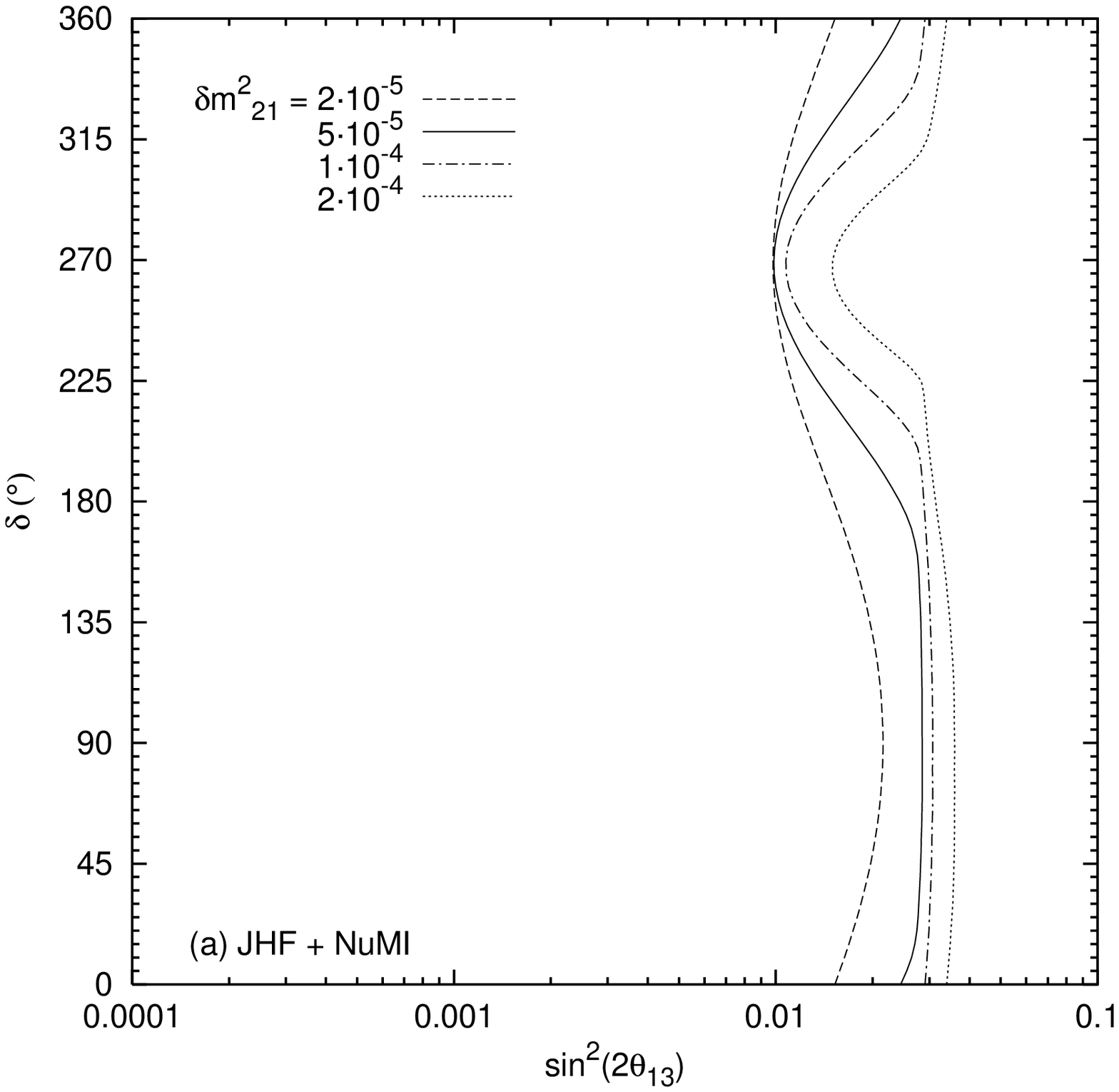,width=8cm,height=12cm}
\psfig{file=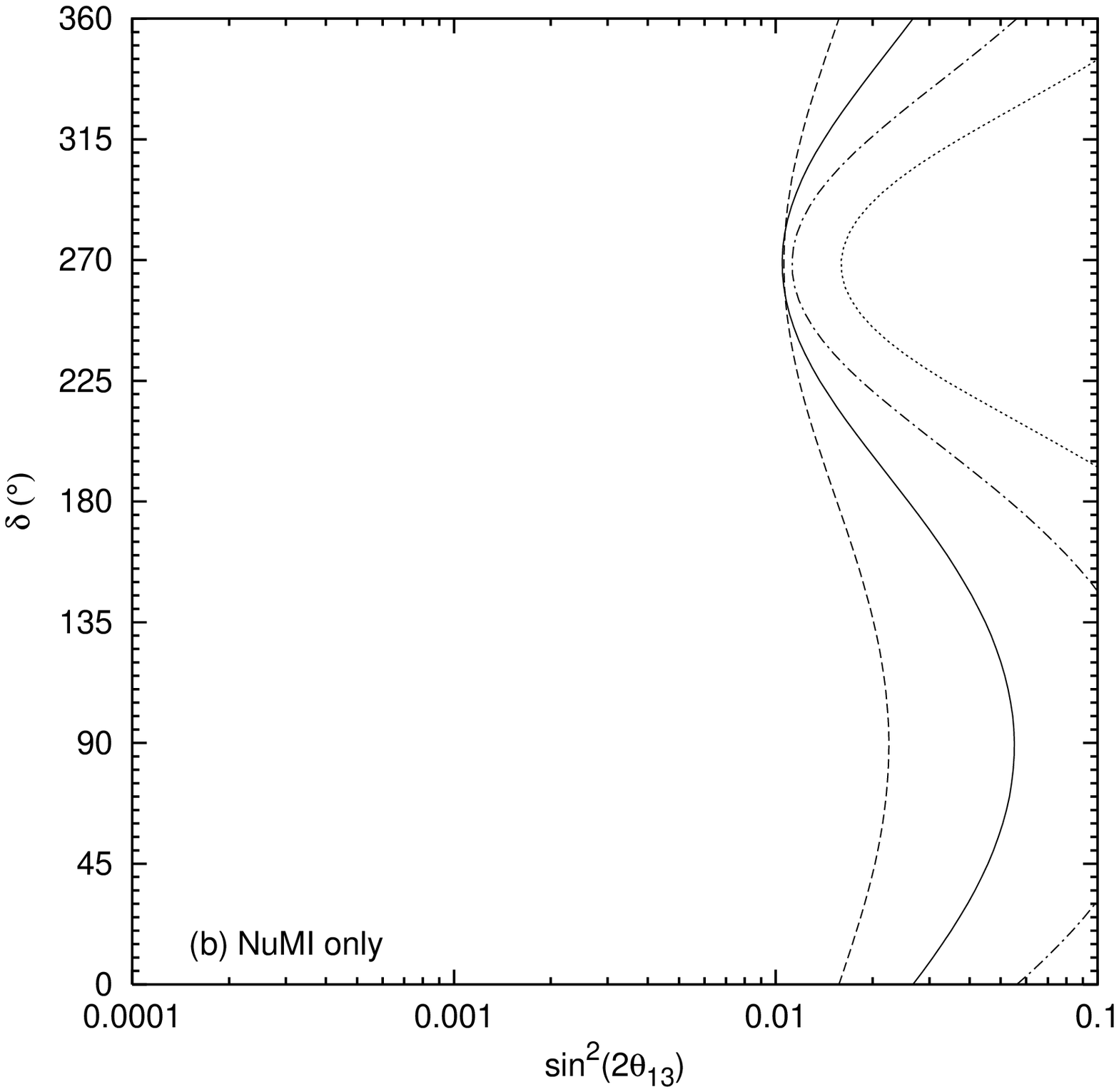,width=8cm,height=12cm}}
\medskip
\caption{Minimum value of $\sin^22\theta_{13}$ for which sgn($\delta
m^2_{31}$) may be determined at $3\sigma$, assuming the true solution has
$\delta m^2_{31} > 0$, using $\nu$ and $\bar\nu$ data from (a) JHF with
$L_1 = 295$~km and NuMI with $L_2 = 890$~km, and (b) only NuMI with
$L_2 = 890$~km, for several values of $\delta m^2_{21}$ (in eV$^2$).
The off-axis angle for the NuMI detector is $\theta_{OA} = 0.74^\circ$.
Other parameters are the same as in Fig.~\ref{fig:discovery}. Results
for $\delta m^2_{31} < 0$ are approximately given by reflecting the
curves about $\delta = 180^\circ$.}
\label{fig:sign3}
\end{figure}

The ability to distinguish the sign of $\delta m^2_{31}$ is greatly
affected by the size of the solar mass scale $\delta m^2_{21}$,
because the predictions for $\delta m^2_{31}>0$ and $\delta
m^2_{31}<0$ overlap more for larger values of $\delta m^2_{21}$. In
Fig.~\ref{fig:sign3}a we show the region in ($\delta,
\sin^22\theta_{13}$) space for which parameters with $\delta
m^2_{31}>0$ can be distinguished from all parameters with $\delta
m^2_{31}<0$ at the $3\sigma$ level for several possible values of
$\delta m^2_{21}$, using combined data from $L_1 = 295$~km and $L_2 =
890$~km, with $\theta_{OA} = 0.74^\circ$ for the second experiment.
With this configuration the $\cos\delta$ terms in the average
probabilities vanish for both experiments and nearly maximal reach for
distinguishing sgn($\delta m^2_{31}$) is achieved. A similar plot using
only data at $L_2 = 890$~km and $\theta_{OA} = 0.74^\circ$ is shown in
Fig.~\ref{fig:sign3}b. We do not show a corresponding plot for $L_1 =
295$~km because the shorter baseline severely inhibits the determination
of sgn($\delta m^2_{31}$). A comparison of the two figures shows that for
$\delta = 270^\circ$ (where the $\delta m^2_{31} > 0$ predictions have
the least overlap with any of those for $\delta m^2_{31} < 0$) the
sensitivity to sgn($\delta m^2_{31}$) is not significantly improved by
adding the data at $L_1$. However, at $\delta = 90^\circ$ the ability
to distinguish sgn($\delta m^2_{31}$) is much less affected by the
value of $\delta m^2_{21}$ when the data at $L_1$ is included. With
data only at $L_2$, sgn($\delta m^2_{31}$) can be determined for
$\sin^22\theta_{13} = 0.1$ when $\delta = 90^\circ$ only for $\delta
m^2_{21} \lsim 8\times10^{-5}$eV$^2$, while with data at $L_1$ and
$L_2$ it can be determined for $\sin^22\theta_{13}$ as low as 0.04 for
$\delta m^2_{21}$ as high as $2\times10^{-4}$~eV$^2$. The
corresponding results for $\delta m^2_{31} < 0$ are approximately
given by reflecting the curves in Fig.~\ref{fig:sign3} about $\delta =
180^\circ$.

We conclude that combining measurements of $\nu_\mu \to \nu_e$ and
$\bar\nu_\mu \to \bar\nu_e$ from two superbeam experiments at
different $L$ results in a much more sensitive test of the sign of
$\delta m^2_{31}$ than one experiment alone, especially for larger
values of the solar mass scale $\delta m^2_{21}$.

The ability to determine sgn($\delta m^2_{31}$) is also affected by
the value of $\theta_{23}$. We found that the $\sin^22\theta_{13}$
reach for determining sgn($\delta m^2_{31}$) at $3\sigma$ varied from
0.02 to 0.04 for $\sin^22\theta_{23} = 0.90$ (compared to 0.03 when
$\theta_{23} = \pi/4$), depending on whether $\delta m^2_{31}$
is positive or negative, and whether $\theta_{23} < \pi/4$ or
$\theta_{23} > \pi/4$.  The sgn($\delta m^2_{31}$) sensitivities for
different possibilities are shown in Table~\ref{tab:sign2}.

\begin{table}[hbtp!]
\begin{center}
\caption{$\sin^22\theta_{13}$ reach for determining the sign of
$\delta m^2_{31}$ at $3\sigma$ using $\nu$ and $\bar\nu$ data from JHF
at 295~km and NuMI at $L_2 = 890$~km with $\theta_{OA} = 0.74^\circ$,
for different combinations of sgn($\delta m^2_{31}$) and $\theta_{23}$,
where $\sin^22\theta_{23} = 0.9$.}
\begin{tabular}{c|cc}
& \multicolumn{2}{c}{$\sin^22\theta_{13}$ reach for sgn($\delta m^2_{31}$)}\\
sgn($\delta m^2_{31}$) & $\sin\theta_{23} = 0.585$
& $\sin\theta_{23} = 0.811$\\
\hline\hline
$+$                    & 0.04  & 0.02\\
$-$                    & 0.03  & 0.03\\
\end{tabular}
\label{tab:sign2}
\end{center}
\end{table}

\subsection{Establishing the existence of $CP$ violation}
\label{sec:CPV}

An important goal of long-baseline experiments is to determine whether
or not $CP$ is violated in the leptonic sector. In order to
unambiguously establish the existence of $CP$ violation, one must be
able to differentiate between ($\delta, \theta_{13}$) and all possible
($\delta^\prime, \theta_{13}^\prime$), where $\delta^\prime = 0^\circ$
or $180^\circ$ and $\theta_{13}^\prime$ can take on any value. For our
$CP$ violation analysis we vary $\sin^22\theta_{13}^\prime$ in 2\%
increments, as was done in the previous section when testing the
sgn($\delta m^2_{13}$) sensitivity.

\begin{figure}
\centering\leavevmode
\psfig{file=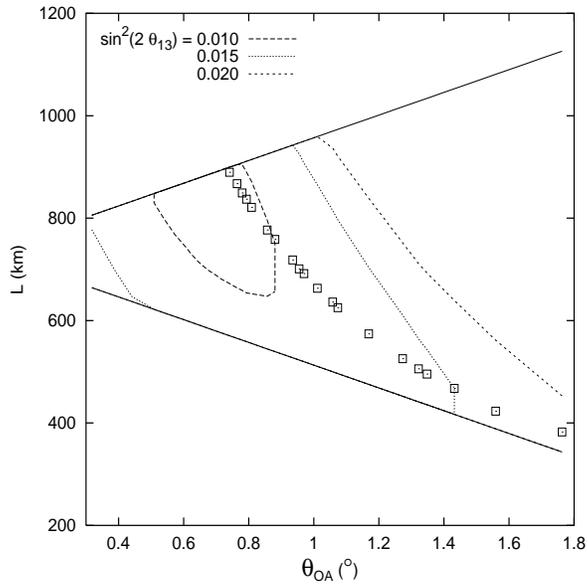,width=8cm,height=12cm}
\medskip
\caption{Contours of $\sin^22\theta_{13}$ reach for distinguishing
$\delta = 90^\circ$ from the $CP$ conserving values $\delta = 0^\circ$
and $180^\circ$ at $3\sigma$ (for the same sgn($\delta m^2_{31}$)),
plotted in the ($\theta_{OA},L_2$) plane, when data from JHF and NuMI
are combined. Other parameters and notation are the same as in
Fig.~\ref{fig:discovery}. Results for $\delta = 270^\circ$ are similar
to those for $\delta = 90^\circ$.}
\label{fig:cpv1}
\end{figure}

Figure~\ref{fig:cpv1} shows contours of $\sin^22\theta_{13}$ reach for
distinguishing $\delta = 90^\circ$ from the $CP$ conserving values
$\delta = 0^\circ$ and $180^\circ$ at $3\sigma$ (with the
same sgn($\delta m^2_{31}$)), plotted in the ($\theta_{OA},L_2$)
plane, assuming $\nu$ and $\bar\nu$ data at both $L_1$ and $L_2$ are
combined. The $CP$ reach in $\sin^22\theta_{13}$ can go as low as 0.01
for $\theta_{OA} \simeq 0.5$ to $0.9^\circ$. Results for $\delta =
270^\circ$ are similar to those for $\delta = 90^\circ$.

\begin{figure}
\centering\leavevmode
\mbox{\psfig{file=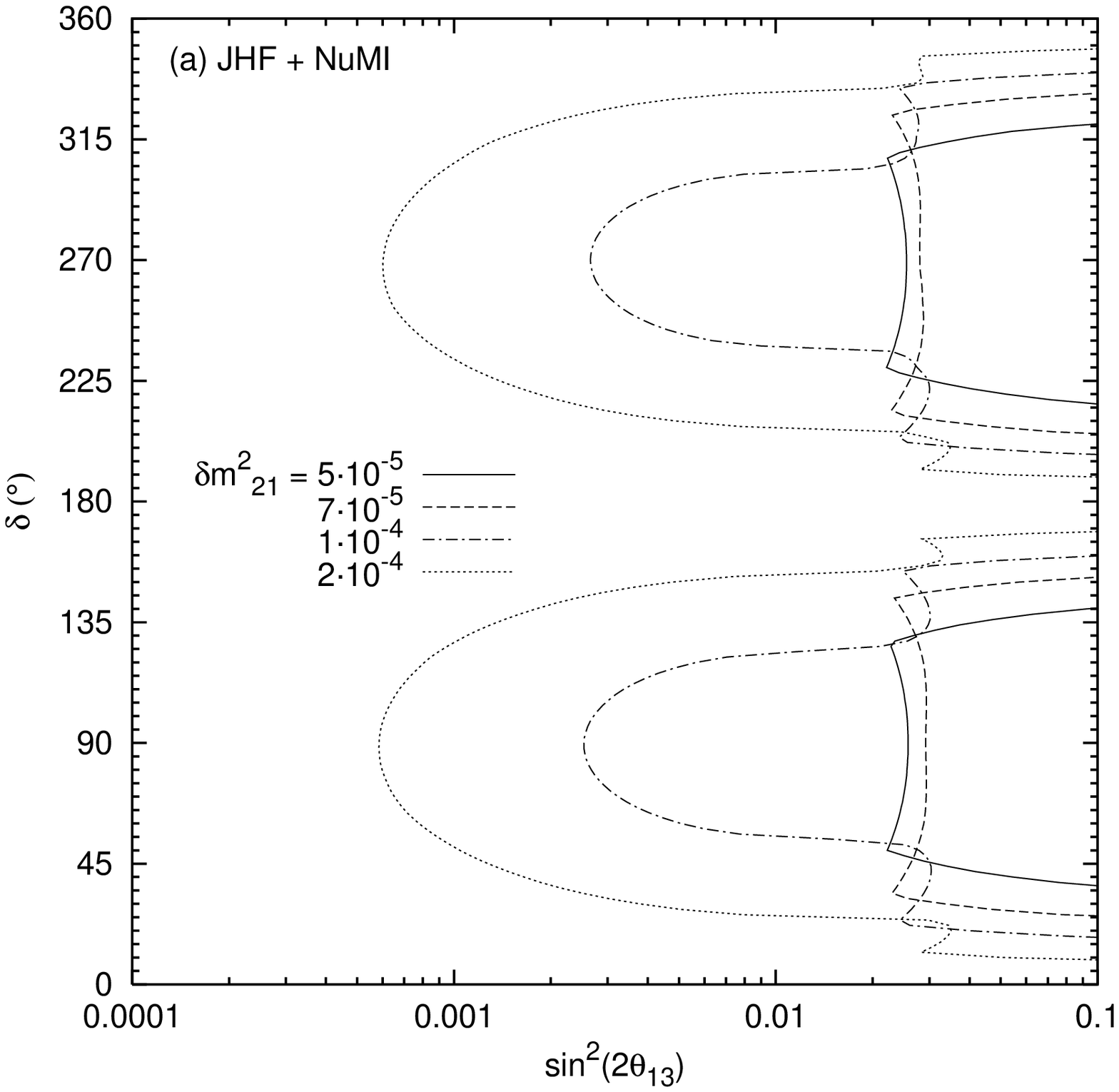,width=8cm,height=12cm}
\psfig{file=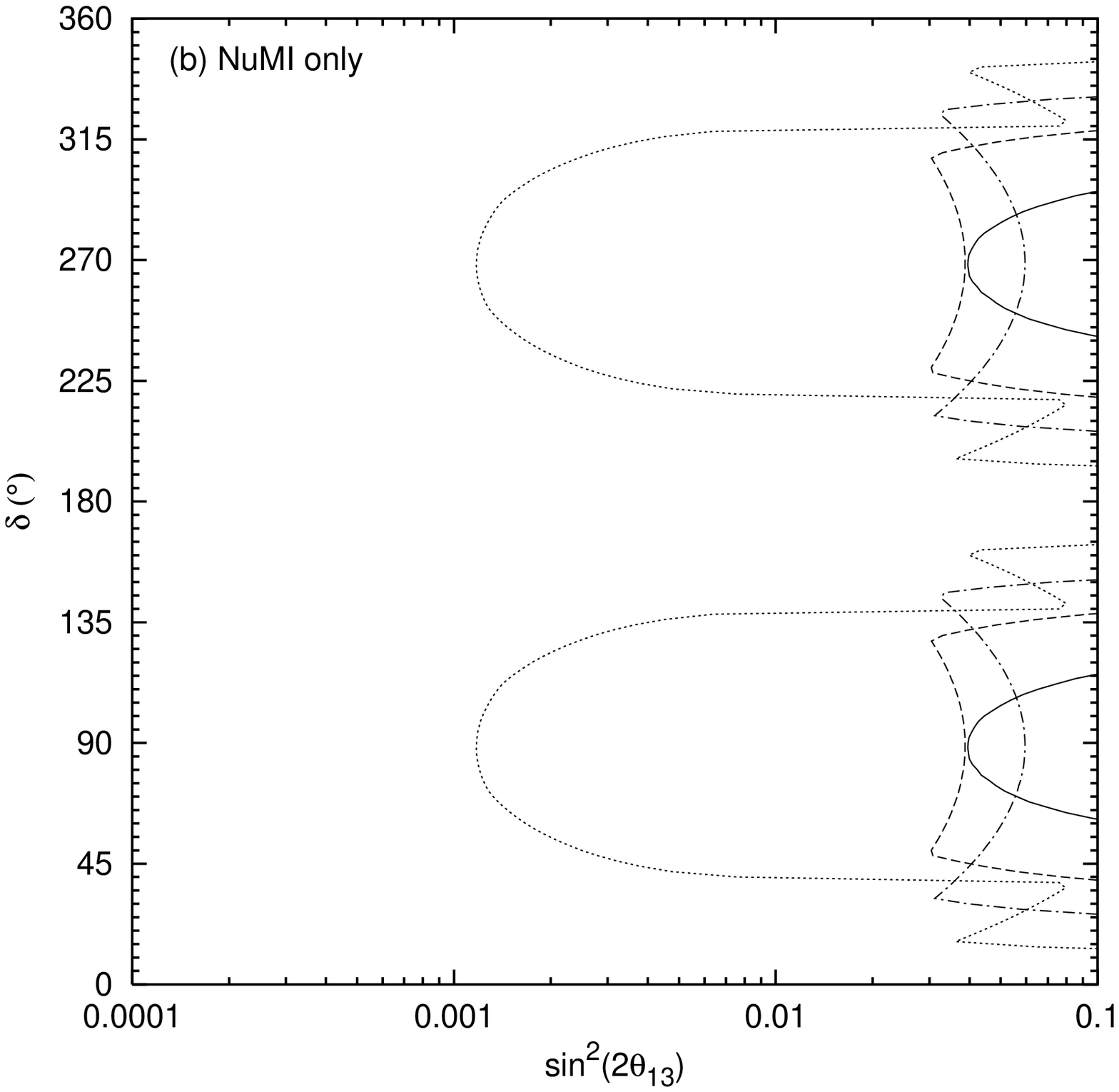,width=8cm,height=12cm}}
\medskip
\caption{Minimum value of $\sin^22\theta_{13}$ versus $CP$ phase
$\delta$ for which $\delta$ can be distinguished from the
$CP$ conserving values $\delta = 0^\circ$ and $180^\circ$ (with either
sign of $\delta m^2_{31}$) at the
$3\sigma$ level when (a) data from JHF and NuMI are combined, and
(b) data from NuMI only are used. The baseline for JHF is $L_1 =
295$~km, while for NuMI $L_2 = 890$~km and $\theta_{OA} = 0.74^\circ$.
The curves are plotted for several values of the solar mass scale
$\delta m^2_{21}$ (in eV$^2$). Other parameters are the same as in
Fig.~\ref{fig:discovery}.}
\label{fig:cpv2}
\end{figure}

Figure~\ref{fig:cpv2} shows the minimum value of $\sin^22\theta_{13}$
for which $\delta$ can be distinguished from all $CP$ conserving
parameter sets with $\delta = 0^\circ$ and $180^\circ$, including
those with the opposite sgn($\delta m^2_{31}$), at the $3\sigma$ level
when $\theta_{OA} = 0.74^\circ$ and $L_2 = 890$~km, for several
different values of $\delta m^2_{21}$. Figure~\ref{fig:cpv2}a shows
the reaches if data from JHF and NuMI are combined, while
Fig.~\ref{fig:cpv2}b shows the reaches if data from NuMI only are used.
For most values of $\delta$, when $\delta m^2_{21}$ is higher the $CP$
effect is increased, and hence $CP$ violation can be detected for
smaller values of $\theta_{13}$. However, there is a possibility that
a $CPV$ solution with one sgn($\delta m^2_{31}$) may not be as easily
distinguishable from a $CPC$ solution with the opposite sgn($\delta
m^2_{31}$); this occurs, e.g., in Fig.~\ref{fig:cpv2}a for $\delta
m^2_{21} = 1\times10^{-4}$~eV$^2$, where the predictions for ($\delta
= 45^\circ$ and $135^\circ$, $\delta m^2_{31} > 0$) are close to those
for ($\delta = 0^\circ$ and $180^\circ$, $\delta m^2_{31} < 0$); in
this case the $CP$ reach for those values of $\delta$ is about the
same for $\delta m^2_{21} = 1\times10^{-4}$~eV$^2$ and $\delta
m^2_{21} = 5\times10^{-5}$~eV$^2$.

We note that if data from only JHF are used (and assuming
$\sin^22\theta_{13} \le 0.1$) no value of the $CP$ phase can be
distinguished at $3\sigma$ from the $CP$ conserving solutions
when $\delta m^2_{21} \lsim 8\times10^{-5}$~eV$^2$, principally
because the intrinsic $CP$ violation due to $\delta$ and the $CP$
violation due to matter have similar magnitudes and it is hard to
disentangle the two effects. For larger values of $\delta m^2_{21}$,
the intrinsic $CP$ effects are larger and $CP$ violation can be
established; e.g., if $\delta m^2_{21} = 1\times10^{-4}$
($2\times10^{-4}$)~eV$^2$, maximal $CP$ violation ($\delta = 90
^\circ$ or $270^\circ$) can be distinguished from $CP$ conservation at
$3\sigma$ for $\sin^22\theta_{13} \gsim 0.006$ ($0.001$). Therefore,
when $\delta m^2_{21} = 1\times10^{-4}$~eV$^2$, most of the $CP$
sensitivity of the combined JHF plus NuMI data results from the JHF
data; for $\delta m^2_{21} = 2\times10^{-4}$~eV$^2$ the two
experiments contribute about equally to the $CP$ sensitivity.

\begin{figure}
\centering\leavevmode
\psfig{file=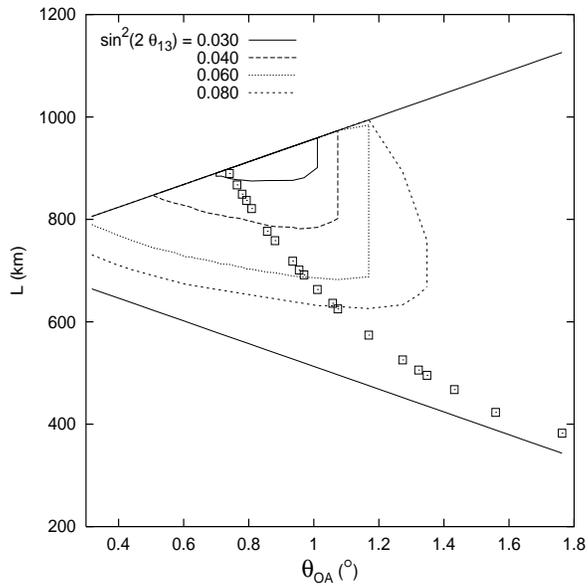,width=8cm,height=12cm}
\medskip
\caption{Contours of $\sin^22\theta_{13}$ reach for resolving the sign
of $\delta m^2_{31}$ at the $3\sigma$ level in the
($\theta_{OA},L_2$) plane when data from JHF and NuMI are
used. The JHF detector is assumed to have baseline $L_1 = 295$~km. Other
parameters and notation are the same as in Fig.~\ref{fig:discovery}.}
\label{fig:sign2}
\end{figure}

The boxes in Figs.~\ref{fig:sign2} and \ref{fig:cpv1} indicate the
values of $L_2$ and $\theta_{OA}$ for which the $\cos\delta$ terms in
the average probabilities vanish for the second experiment. As
indicated in the figures, these detector positions are good for both
distinguishing sgn($\delta m^2_{31}$) (see Fig.~\ref{fig:sign2}) and
for establishing the existence of $CP$ violation (see
Fig.~\ref{fig:cpv1}), especially for larger values of $L_2$. A good
compromise occurs at $\theta_{OA} \simeq 0.74^\circ$ with $L_2 \simeq
890$~km. In Ref.~\cite{barenboim} it was shown that similar values for
$\theta_{OA}$ and $L_2$ using the NuMI off-axis beam gave a favorable
figure-of-merit for the signal to background ratio; our analysis shows
that such an off-axis angle and baseline is also very good for
distinguishing sgn($\delta m^2_{31}$) and establishing $CP$ violation,
when combined with superbeam data at $L_1 = 295$~km.

\subsection{Resolving the ($\delta, \pi - \delta$) ambiguity}
\label{sec:delta}

If $L_2 \simeq 890$~km and $\theta_{OA} \simeq 0.74^\circ$ are chosen
for the location of the second experiment, as suggested in the
previous section, then both the first and second experiments are
effectively measuring $\sin\delta$, and it is impossible to resolve
the ($\delta,\pi-\delta$) ambiguity. Different values of $L_2$ and
$\theta_{OA}$ would be needed to distinguish $\delta$ from $\pi -
\delta$.

\begin{figure}
\centering\leavevmode
\psfig{file=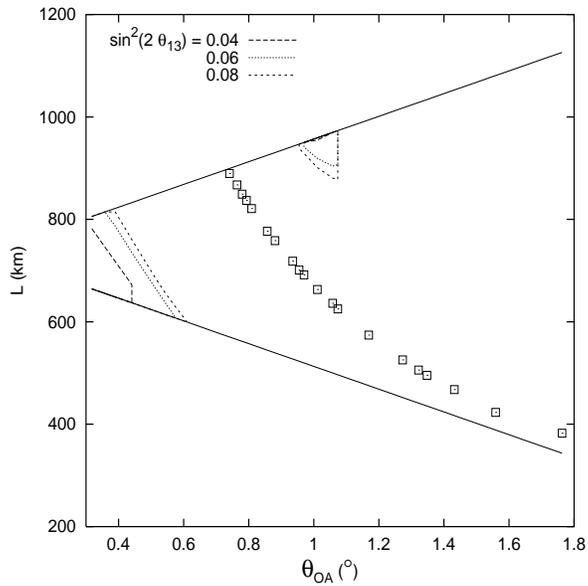,width=8cm,height=12cm}
\medskip
\caption{Contours of $\sin^22\theta_{13}$ reach for distinguishing
$\delta = 0^\circ$ from $\delta = 180^\circ$ (the $(\delta, \pi -
\delta)$ ambiguity) at the $2\sigma$ level, plotted in the
($\theta_{OA},L_2$) plane, when data from JHF and NuMI are combined.
Other parameters and notation are the same as in Fig.~\ref{fig:discovery}.}
\label{fig:delta}
\end{figure}

Figure~\ref{fig:delta} shows contours (in the space of possible $L_2$
and $\theta_{OA}$) for the minimum value of $\sin^22\theta_{13}$
needed to distinguish $\delta = 0^\circ$ from $\delta = 180^\circ$ at
the $2\sigma$ level using $\nu$ and $\bar\nu$ data from $L_1$ and
$L_2$ (it is not possible to distinguish $\delta = 0^\circ$ from
$\delta = 180^\circ$ at the $3\sigma$ level for any value of
$\sin^22\theta_{13}\le0.1$). Two choices are possible: one with
$\theta_{OA} \lsim 0.3$-$0.5^\circ$ and $L_2 \simeq 650$-$775$~km, and
another near $\theta_{OA} \simeq 1.0^\circ$ with $L_2 \simeq 950$~km.
The former choice does not do well in distingishing sgn($\delta
m^2_{31}$), while the latter choice is nearly optimal for sgn($\delta
m^2_{31}$) sensitivity but significantly worse for $CP$ violation
sensitivity. Thus the ability to also resolve the ($\delta, \pi -
\delta$) ambiguity is rather poor, and comes at the expense of $CPV$
sensitivity.

\subsection{Resolving the ($\theta_{23},\pi/2-\theta_{23}$) ambiguity}
\label{sec:theta}

If $\theta_{23} \ne \pi/4$, there is an additional ambiguity between
$\theta_{23}$ and $\pi/2-\theta_{23}$. This ambiguity gives two
solutions for $\sin^22\theta_{13}$ whose ratio differs by a factor of
approximately $\tan^2\theta_{23}$, which can be as large as 2 if
$\sin^22\theta_{23} = 0.9$~\cite{bmw}. Assuming $L_1 = 295$~km for the
first experiment, we could not find any experimental configuration of
$L_2$ and $\theta_{OA}$ for the second experiment that could resolve
the ($\theta_{23}, \pi/2-\theta_{23}$) ambiguity for
$\sin^22\theta_{13} \le 0.1$ at even the $1\sigma$ level for the
entire range of detector sizes and source powers listed in
Table~\ref{tab:sign1}. Therefore we conclude that superbeams are not
effective at resolving the ($\theta_{23}, \pi/2-\theta_{23}$)
ambiguity using $\nu_e$ and $\bar\nu_e$ appearance data. Since the
approximate oscillation probability for $\nu_e \to \nu_\tau$ is given
by the interchanges $\sin\theta_{23} \leftrightarrow \cos\theta_{23}$
and $\delta \to - \delta$ in the expression for the $\nu_\mu \to
\nu_e$ probability, a neutrino factory combined with detectors having
tau neutrino detection capability provides a means for resolving the
($\theta_{23}, \pi/2-\theta_{23}$) ambiguity~\cite{bmw}.
Another possibility is to measure survival of $\bar\nu_e$'s from a
reactor, which to leading order is sensitive to $\sin^22\theta_{13}$
but not $\theta_{23}$~\cite{degouvea,minakata3}.

\subsection{Dependence on $|\delta m^2_{31}|$}
\label{sec:dependence}

The foregoing analysis assumed $|\delta m^2_{31}| =
3\times10^{-3}$~eV$^2$. If the true value differs from this, then to
sit on the peak (where the $\cos\delta$ terms vanish) requires tuning
the beam energy and baseline according to the measured value of
$|\delta m^2_{31}|$. JHF has the capability of varying the average
$E_\nu$ from 0.4~GeV to 1.0~GeV, which would correspond to realizing
the peak condition for $|\delta m^2_{31}| =
1.6$-$4.0\times10^{-3}$~eV$^2$~\cite{jhfsk}. In principle, NuMI can
vary both $L_2$ and $\theta_{OA}$ to be on the peak.  If $|\delta
m^2_{31}| < 3\times10^{-3}$~eV$^2$, then the best sensitivity to
sgn($\delta m^2_{31}$) is obtained for larger $\theta_{OA}$ and longer
distances (the larger angle makes $E_\nu$ smaller while the longer
distance enhances the matter effect), and the sensitivity is reduced
(since the matter effect is smaller for smaller $\delta
m^2_{31}$). The $CP$ violation sensitivity is also reduced, although
not as significantly. For larger values of $|\delta m^2_{31}|$ the
sensitivity to sgn($\delta m^2_{31}$) is better, with $CP$
violation sensitivity about the same.

The tuning of the experiments to the peak (where the $\cos\delta$
terms in the average probabilities vanish) requires knowledge of
$|\delta m^2_{31}|$ before the experimental design is finalized. The
values of $|\delta m^2_{31}|$ and $\theta_{23}$ will be well-measured
in the survival channel $\nu_\mu \to \nu_\mu$ measurements that would
run somewhat before or concurrently with the appearance measurements
being discussed here, but of course this information may not be
available when the configurations for the off-axis experiments are
chosen. If $|\delta m^2_{31}|$ is known to 10\% at $3\sigma$ (the
expected sensitivity of MINOS), then the sensitivities to sgn($\delta
m^2_{31}$) and $CP$ violation are not greatly affected by baselines
that are slightly off-peak. If the baselines and neutrino energies for
the superbeam experiments must be chosen before a 10\% measurement of
$|\delta m^2_{31}|$ can be made, a loss of sensitivity to sgn($\delta
m^2_{31}$) could result by not being on the peak. For example, if the
experiments are designed for $|\delta m^2_{31}| =
3\times10^{-3}$~eV$^2$ but in fact $|\delta m^2_{31}| =
2.5\times10^{-3}$~eV$^2$, the $3\sigma$ sgn($\delta m^2_{31}$) reach
is less ($\sin^22\theta_{13} = 0.04$, compared to 0.03 for $|\delta
m^2_{31}| = 3\times10^{-3}$~eV$^2$). If $|\delta m^2_{31}|$ is
actually $2\times10^{-3}$~eV$^2$, the $3\sigma$ sgn($\delta
m^2_{31}$) reach extends only down to $\sin^22\theta_{13} \simeq
0.075$, just a little below the CHOOZ bound.

Since the sgn($\delta m^2_{31}$) determination has the worst reach in
$\sin^22\theta_{13}$ (compared to the discovery reach and the $CPV$
sensitivity), and since not knowing sgn($\delta m^2_{31}$) can induce
a $CPV/CPC$ ambiguity, the measurement of sgn($\delta m^2_{31}$) is
crucial. If $|\delta m^2_{31}|$ is not known precisely, then the exact
peak position is not known, and an off-axis angle and baseline should
be chosen that will give a reasonable reach for sgn($\delta m^2_{31}$)
over as much of the allowed range of $|\delta m^2_{31}|$ as
possible. For example, $\theta_{OA} = 0.85^\circ$-$0.90^\circ$ and $L
\simeq 930$~km gives a sgn($\delta m^2_{31}$) reach that is fairly
good for the range $|\delta m^2_{31}| = 2\times10^{-3}$~eV$^2$ to
$4\times10^{-3}$~eV$^2$. The reach for
sgn($\delta m^2_{31}$) is farthest from optimal at the extremes
($\sin^22\theta_{13}=0.06$ versus the best reach of 0.05 when $|\delta
m^2_{31}| = 2\times10^{-3}$~eV$^2$ and 0.03 versus the best reach of
0.02 when $|\delta m^2_{31}| = 4\times10^{-3}$~eV$^2$). But the
$CPV$ reach remains at least as good as the sgn($\delta m^2_{31}$)
reach for this range of $|\delta m^2_{31}|$.

\section{Summary}
\label{sec:summary}

We summarize the important points of our paper as follows:

\begin{enumerate}

\item[(i)] Two superbeam experiments at different baselines, each
measuring $\nu_\mu \to \nu_e$ and $\bar\nu_\mu \to \bar\nu_e$
appearance, are significantly better at resolving the sgn($\delta
m^2_{31}$) ambiguity than one experiment alone. Using beams from a
4.0~MW JHF with a 22.5~kt detector $2^\circ$ off axis at 295~km and a
1.6~MW NuMI with a 20~kt detector $0.7$-$1.0^\circ$ off axis at
$875$-$950$~km, sgn($\delta m^2_{31}$) can be determined for
$\sin^22\theta_{13} \gsim 0.03$ if $\delta m^2_{31} =
3\times10^{-3}$~eV$^2$. Sensitivities for other beam powers and
detector sizes are given in Table~\ref{tab:sign1}.

\item[(ii)] For the most favorable cases, a higher value for the solar
oscillation scale $\delta m^2_{21}$ does not greatly change the
sensitivity to sgn($\delta m^2_{31}$) when $\nu$ and $\bar\nu$ data
from two different baselines are combined (unlike the single baseline
case, where the ability to determine sgn($\delta m^2_{31}$) is
significantly worse for $\delta m^2_{21} \gsim 5\times10^{-5}$~eV$^2$).

\item[(iii)] Running both experiments at the oscillation peaks, such
that the $\cos\delta$ terms in the average probabilities vanish,
provides good sensitivity to both sgn($\delta m^2_{31}$) and to $CP$
violation. On the other hand, the ability to resolve the ($\delta,
\pi-\delta$) ambiguity is lost, and the ($\theta_{23},
\pi/2-\theta_{23}$) ambiguity is not resolved for any experimental
arrangement considered. However, the ($\delta, \pi-\delta$)
and ($\theta_{23}, \pi/2-\theta_{23}$) ambiguities do not
substantially affect the ability to determine whether or not $CP$ is
violated (although the latter ambiguity could affect the inferred
value of $\theta_{13}$ by as much as a factor of 2).

\item[(iv)] Since running at or near the oscillation peaks is
favorable, knowledge of $|\delta m^2_{31}|$ to about 10\% (from MINOS)
before these experiments are run would be advantageous. If $|\delta
m^2_{31}|$ is not known that precisely in advance, then the detector
off-axis angle and baseline can still be chosen to give fairly good
(though not optimal) sensitivities to sgn($\delta m^2_{31}$) and $CP$
violation.

\end{enumerate}

We conclude that superbeam experiments at different baselines may
greatly improve the prospects for determining the neutrino mass
ordering in the three-neutrino model. Since a good compromise between
determining sgn($\delta m^2_{31}$) and establishing the existence of
$CP$ violation is obtained when both experiments are tuned so that the
$\cos\delta$ terms in the average probabilities approximately vanish,
knowledge of $|\delta m^2_{31}|$ would be helpful for the optimal
design for the experiments.


\section*{Acknowledgments}

We thank A. Para for information on the NuMI off-axis beams, and
A. Para and D. Harris for helpful discussions. This research was
supported in part by the U.S.~Department of Energy under Grants
No.~DE-FG02-95ER40896, No.~DE-FG02-01ER41155 and
No.~DE-FG02-91ER40676, and in part by the University of Wisconsin
Research Committee with funds granted by the Wisconsin Alumni Research
Foundation.

\clearpage

%

%

%

%

%

%

%

\end{document}